\documentclass[
    referee,       
    pdflatex,
    sn-chicago,
    iicol
]{sn-jnl}
\usepackage[utf8]{inputenc}

\usepackage[T1]{fontenc}
\usepackage{xcolor}
\usepackage{enumitem}  
\usepackage{graphicx}
\usepackage{amsmath,amssymb}
\usepackage{booktabs}

\usepackage{array}        
   
\usepackage{hyperref}

\begin{document}

\title{Enhanced Consistency Bi-directional GAN
(CBiGAN) for Malware Anomaly Detection}

\author*[1]{\fnm{Thesath} \sur{Wijayasiri}}
\email{bopearachchigethesathguwantha@stengg.com}

\author[1]{\fnm{Fok} \sur{Kar Wai}}
\author[1]{\fnm{Vrizlynn L. L.}   \sur{Thing}}

\affil[1]{\orgdiv{Cybersecurity Strategic Technology Centre},
          \orgname{ST Engineering},
          \city{Singapore}}

\abstract{
Static malware analysis remains a core technique in cybersecurity due to its ability to assess potentially malicious software without execution. Nevertheless, many existing static approaches rely on handcrafted features or curated datasets that may not generalize well to evolving malware distributions. In this work, we investigate an alternative representation that operates directly on raw binary content. Executable files are transformed into visual encodings that preserve local structural relationships, enabling the use of deep learning models without requiring semantic disassembly or dynamic behavior profiling.

This study explores the use of a Consistency Bi-directional Generative Adversarial Network (CBiGAN) as an anomaly detection framework rather than as a generative model. The method enforces consistency between latent encodings and reconstructions, allowing deviations from learned benign structure to be quantified through reconstruction discrepancies. Importantly, the approach does not introduce a new generative architecture, instead, it evaluates how consistency based generative modeling can be applied at scale to heterogeneous malware data.

The proposed framework is evaluated across multiple datasets comprising both Portable Executable (PE) and Object Linking and Embedding (OLE) files, including a large self-collected corpus spanning 214 malware families. Results demonstrate stable detection performance in terms of Area Under the Curve (AUC) while maintaining a unified and computationally lightweight processing pipeline. These findings suggest that consistency based generative modeling provides a practical and scalable direction for malware anomaly detection across diverse file formats and threat families.
}


\maketitle

\section*{List of abbreviations}
\begin{description}[style=multiline,leftmargin=2.8cm]
  \item[AE] Autoencoder
  \item[AUC] Area Under the (ROC) Curve
  \item[BiGAN] Bidirectional Generative Adversarial Network
  \item[CBiGAN] Consistency Bidirectional Generative Adversarial Network
  \item[CNN] Convolutional Neural Network
  \item[EMA] Exponential Moving Average
  \item[GAN] Generative Adversarial Network
  \item[OLE] Object Linking and Embedding
  \item[PCA] Principal Component Analysis
  \item[PDF] Portable Document Format
  \item[PE] Portable Executable
  \item[RGB] Red Green Blue colour model
  \item[ROC] Receiver Operating Characteristic
\end{description}

\section{Introduction}
Malware detection remains a major challenge in cybersecurity, serving as a primary defense against unauthorized access, data exfiltration, and system compromise. Among existing detection strategies, static analysis plays a crucial role due to its ability to examine software without execution, thereby avoiding the risks associated with running potentially malicious code. This characteristic makes static analysis particularly suitable for large scale and high security environments. However, this advantage often comes at the cost of increased analysis time and reduced adaptability when facing rapidly evolving malware variants.

A limitation of many static malware detection systems is their reliance on biased, incomplete, or outdated datasets. Such datasets frequently fail to capture the diversity and evolution of contemporary malware. This results in a diminished detection capability when models encounter previously unseen threats. As malware authors continue to employ obfuscation, packing, and polymorphism to evade signature based defenses, there is an increasing need for approaches that generalize beyond known patterns and instead identify deviations from benign behavior.

One research direction that has attracted attention is the representation of malware binaries as images. By converting binary content into visual form, global structural and texture patterns within executables can be preserved while avoiding reliance on semantic disassembly or dynamic execution traces. This idea was initially introduced by Nataraj et al. \cite{nataraj2011malware}, who demonstrated that visual encodings of binaries enable the application of image based machine learning techniques to malware detection. Unlike feature sets derived from localized properties such as API calls or opcode sequences, these visual representations emphasize global characteristics of binaries, capturing layout and structural regularities that may persist even under obfuscation.

Building on this, deep learning models have been widely adopted to analyze malware images. At the same time, Generative Adversarial Networks (GANs) have shown strong capability in modeling complex data distributions in image domains. While GANs are commonly associated with data synthesis, their utility extends beyond generation to tasks such as representation learning and anomaly detection. In particular, bidirectional GAN variants introduce an encoder alongside the generator, enabling simultaneous mapping between data and latent space. Sabuhi et al. \cite{sabuhi2021applications} provides a comprehensive overview of GAN architectures and their applicability across different data modalities. This highlights bidirectional models as especially effective for complex and texture rich imagery.

Motivated by these observations, this work investigates the use of the Consistency Bi-directional Generative Adversarial Network (CBiGAN) \cite{carrara2021combining} as an anomaly detection framework for malware analysis. Rather than treating the model as a generative tool, we leverage its latent space consistency mechanism to quantify deviations between encoded representations and reconstructions. The suitability of CBiGAN for this task is informed by its demonstrated performance on texture oriented anomaly detection benchmarks, such as the MVTec dataset, suggesting potential effectiveness in capturing subtle structural irregularities present in malware imagery.

The proposed framework applies this consistency based modeling approach to both Portable Executable (PE) and Object Linking and Embedding (OLE) files, with training performed exclusively on benign samples. To better accommodate the complexity of malware images, the base encoder and generator of the CBiGAN is replaced with established deep convolutional encoder architectures, including ResNet, DenseNet, and Inception variants. The impact of encoder selection is systematically evaluated to assess its influence on detection performance and generalization.

Experimental evaluation is conducted using a diverse collection of datasets, including 6,330 benign PE files, 10,820 malicious executables spanning 214 malware families, and a large set of benign and malicious PDF documents from the Contagio dataset \cite{mila2013dataset}. In addition, transferability is examined using selected classes from the Microsoft Malware Challenge dataset. Results demonstrate that the proposed approach achieves strong detection performance in terms of Area Under the Curve (AUC) while maintaining a simple and unified processing pipeline.

Recent malware anomaly detection research has increasingly explored self supervised representation learning, diffusion based generative models, and flow based density estimation techniques. While these approaches often achieve strong results in constrained experimental settings, they frequently incur high computational cost and may exhibit sensitivity to feature leakage or dataset specific artifacts. In contrast, this work positions CBiGAN based consistency learning as a lightweight and scalable alternative, emphasizing robustness to data heterogeneity and transferability across malware families rather than maximizing performance on a single benchmark.

\section{Related Work}

Static malware analysis has attracted increasing research interest due to its ability to inspect executables without execution, thereby avoiding the risks associated with dynamic analysis. Prior work has explored both local feature representations, such as opcode sequences and API calls, as well as global feature representations derived from entire byte streams \cite{wu2021survey,sihwail2018survey,nataraj2011malware,ngo2020survey,bensaoud2024survey}. Global representations are particularly attractive in adversarial settings where semantic parsing may be unreliable due to packing, obfuscation, or polymorphism. In this work, we adopt a one class learning method, also referred to as anomaly detection, which models benign software behavior and identifies deviations indicative of malicious activity.

\subsection{One Class Classification for Malware Detection}

One class classification approaches train models exclusively on benign data and identify anomalies based on deviations from the learned distribution. Classical techniques include One Class - Support Vector Machines (OC-SVM), which learn a separating boundary in high dimensional feature space, and Isolation Forests, which identify anomalies through recursive partitioning. In the malware domain, one class methods have been widely adopted to address class imbalance and the open set nature of real world threats, where malicious samples may be scarce, noisy, or continuously evolving.

While traditional one class models can be effective for structured or low dimensional feature spaces, they often struggle to capture the complexity of modern malware. This limitation has motivated the use of deep learning based anomaly detection methods, including autoencoders and generative models, which are capable of learning richer representations. However, autoencoder based approaches may over generalize, reconstructing anomalous inputs too accurately when trained on diverse datasets, thereby reducing detection sensitivity.

\subsection{Generative Adversarial Networks for Anomaly Detection}

\begin{figure}
    \centering
    \includegraphics[width=1\linewidth]{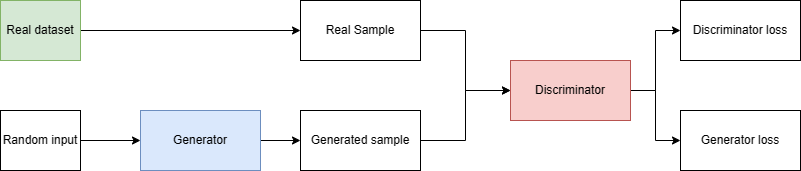}
    \caption{Generative Adversarial Network}
    \label{GAN}
\end{figure}

Generative Adversarial Networks (GANs), introduced by Goodfellow et al. \cite{goodfellow2014gan}, consist of a generator and a discriminator trained in an adversarial setting to model complex data distributions as shown in figure \ref{GAN}. The generator learns to synthesize samples from a latent distribution, while the discriminator attempts to distinguish generated samples from real data. Through this adversarial process, GANs can capture intricate structures in high-dimensional data.

\begin{figure}
    \centering
    \includegraphics[width=0.75\linewidth]{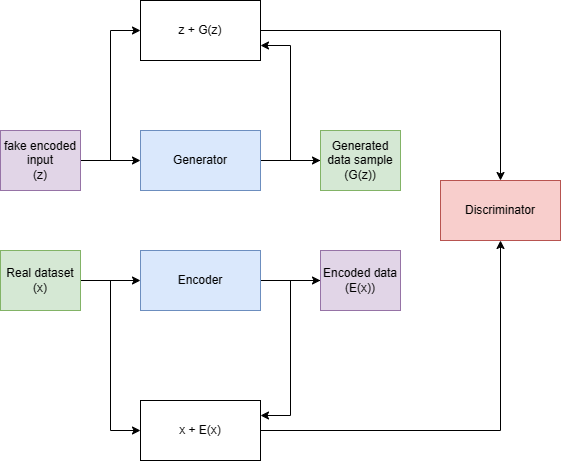}
    \caption{Bi-directional GAN}
    \label{BiGAN}
    \vspace{-3mm}
\end{figure}

Beyond data synthesis, GANs have been increasingly applied to anomaly detection by exploiting reconstruction errors or discriminator based feature discrepancies. Bidirectional GAN variants, such as BiGAN \cite{donahue2016adversarial} as shown in figure \ref{BiGAN}, extend the original framework by introducing an encoder that maps data samples into latent space, enabling bidirectional translation between data and latent representations.

\subsection{Consistency Bi-directional GAN (CBiGAN)}

\begin{figure*}
    \centering
    \includegraphics[width=\textwidth]{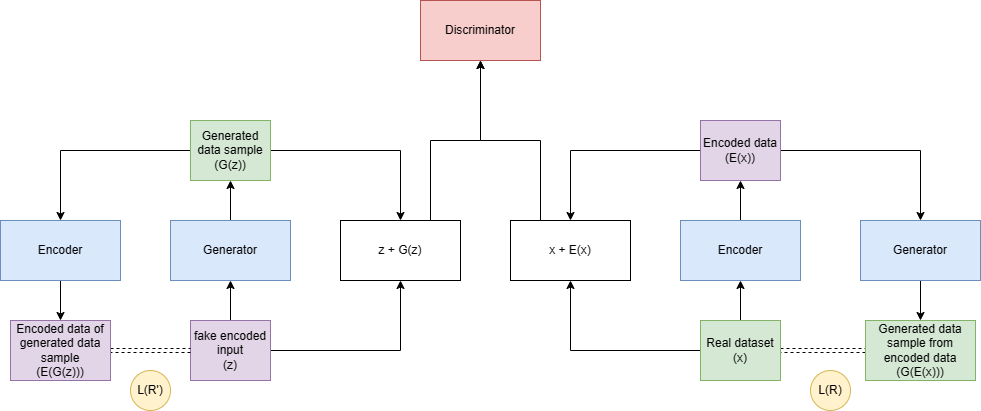}
    \caption{Consistency Bi-GAN}
    \label{CBiGAN}
\end{figure*}

The Consistency Bi-directional GAN (CBiGAN) as shown in figure \ref{CBiGAN} \cite{carrara2021combining} further extends the BiGAN framework by introducing consistency constraints between the encoder and generator as a regularization mechanism.

Unlike traditional GANs that emphasize sample generation, CBiGAN is explicitly designed for anomaly detection. During inference, anomaly scores are computed as a combination of pixel level reconstruction error and feature level discrepancies extracted from the discriminator. Samples that deviate from the benign training distribution exhibit increased reconstruction inconsistency, making the framework particularly suitable for texture oriented anomaly detection tasks.

\subsection{Malware Visualization and Image Based Analysis}

Representing malware binaries as images has emerged as an effective strategy for applying computer vision techniques to static malware analysis. Nataraj et al. \cite{nataraj2011malware} demonstrated that converting raw byte sequences into grayscale images preserves structural and texture patterns that are often shared within malware families. Subsequent studies have explored alternative visualization strategies, including color encodings, different pixel mapping schemes, and deep convolutional architectures for classification and analysis.

Malware visualization methods emphasize global structural characteristics rather than localized semantic features, which makes them more robust to obfuscation and code reordering. Space filling curves, such as the Hilbert curve \cite{vu2019hit4mal}, have been proposed to preserve spatial locality when mapping high dimensional byte streams into two dimensional representations. Color mapping strategies further enrich these representations by encoding additional structural information across channels. While these techniques have been widely studied for malware classification, their use in anomaly detection remains comparatively underexplored.

\subsection{Recent Self-Supervised and Generative Anomaly Detection}

Recent work in anomaly detection has increasingly focused on self supervised representation learning and probabilistic generative modeling. Approaches based on masked autoencoders, contrastive learning frameworks (e.g., SimCLR, DINO), and reconstruction based pretraining have demonstrated strong performance in visual anomaly detection by learning dense semantic representations of normal data. At the same time, diffusion based models and flow based likelihood estimation techniques have been investigated due to their expressive modeling capacity and principled density estimation.

Despite their success in controlled settings, these methods often incur substantial computational cost and may exhibit sensitivity to feature leakage or dataset specific artifacts. Moreover, their reliance on semantic consistency can limit transferability when applied to highly textured or heterogeneous data such as malware binaries represented as images. These limitations motivate the exploration of alternative anomaly detection methods that emphasize structural consistency over explicit density estimation.

In contrast to these methods, our approach does not rely on dense semantic representations or explicit likelihood modeling, but instead models consistency between learned latent spaces, which empirically demonstrates improved transferability across malware families and file formats.

\subsection{Positioning of This Work}

Within the context of malware image based anomaly detection, this work differs from prior approaches such as Shaukat et al. \cite{shaukat2024detection} in both representation and pipeline design. While their framework employs deep convolutional networks as feature extractors followed by dimensionality reduction and a separate OC-SVM classifier, our approach adopts a unified generative consistency model without explicit feature selection or postprocessing stages. Additionally, whereas many existing studies rely heavily on the MalIMG dataset, which reflects malware characteristics from over a decade ago, this work evaluates performance on more recent and diverse datasets spanning 214 malware families.

By combining lightweight image preprocessing, locality preserving visualization, and consistency based generative modeling, the proposed framework emphasizes scalability, reproducibility, and robustness to data heterogeneity. This design choice reflects an emphasis on transferability and practical applicability rather than highly specialized, multistage detection pipelines.

\section{Proposed Methodology}

Our methodology is designed around the analysis of global structural features while deliberately minimizing preprocessing complexity. This design choice reflects an emphasis on practicality and reproducibility, aiming to construct a streamlined pipeline that could plausibly operate in constrained environments such as edge or local deployment scenarios. By prioritizing global representations, the approach captures broad contextual information from binaries, which is particularly valuable when dealing with heterogeneous and obfuscated malware. Reducing preprocessing overhead further contributes to faster execution and lowers the risk of error propagation, both of which are important considerations for large scale or time sensitive detection settings.

Importantly, the proposed CBiGAN formulation is used as a \emph{consistency enforcing mechanism} rather than as a generative sample tool, aligning it more closely with recent reconstruction based anomaly detection methods.

\begin{figure}
    \centering
    \includegraphics[width=1\linewidth]{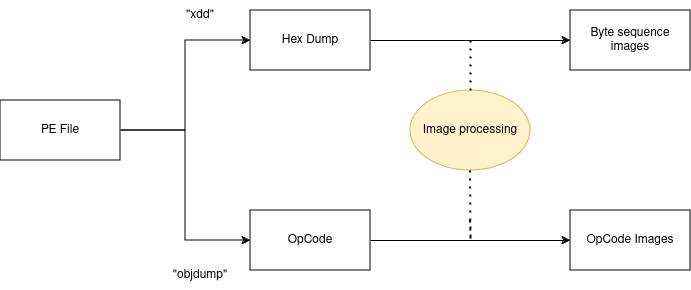}
    \caption{PE to image conversion process}
    \label{PEimageconversion}
\end{figure}

\begin{figure}
    \centering
    \includegraphics[width=1\linewidth]{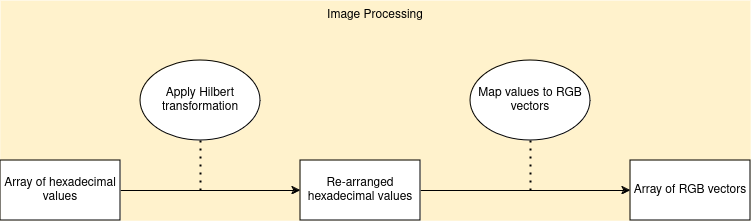}
    \caption{Image Conversion}
    \label{Overallimageconversion}
    \vspace{-3mm}
\end{figure}

Figure \ref{PEimageconversion} illustrates the overall pipeline for converting PE files into images. The same procedure is applied to OLE files. Each file is first converted into an array of hexadecimal values, which then undergo an image processing stage to produce the final visual representation. Figure \ref{Overallimageconversion} details this image processing step, where hexadecimal values are rearranged using a Hilbert space filling curve and subsequently mapped to RGB vectors. The resulting images are used to train a modified version of the CBiGAN proposed by Carrara et al.~\cite{carrara2021combining}, in which a deep convolutional network replaces the original encoder.

The model operates as a one class classifier and is trained exclusively on images derived from benign software. During inference, samples originating from malicious files are expected to produce higher reconstruction discrepancies, resulting in elevated anomaly scores.

\subsection{Datasets}

\subsubsection{Primary Dataset}

The primary experiment uses a self-collected dataset comprising 6,330 benign Portable Executable (PE) files after cleaning. These files were obtained using a web crawler and verified as benign via VirusTotal. Hashing was employed to eliminate duplicate samples, and files smaller than 2\,MB or larger than 50\,MB were excluded to align size distributions with the malicious dataset. The benign data were split into training, validation, and test sets using a 60/20/20 ratio, yielding 3,798 training samples and 1,266 samples each for validation and testing.

The malicious dataset consists of 10,820 PE samples spanning 214 malware families, sourced from MalwareBazaar~\cite{malwarebazaar}. Samples were selected based on upload date, ranging from 2008 to 2023, with more than 80\% originating between 2018 and 2023. All samples were revalidated using VirusTotal, deduplicated via hashing, and filtered to retain only files within the 2\,MB to 50\,MB size range.

\begin{figure}
    \centering
    \includegraphics[width=1\linewidth]{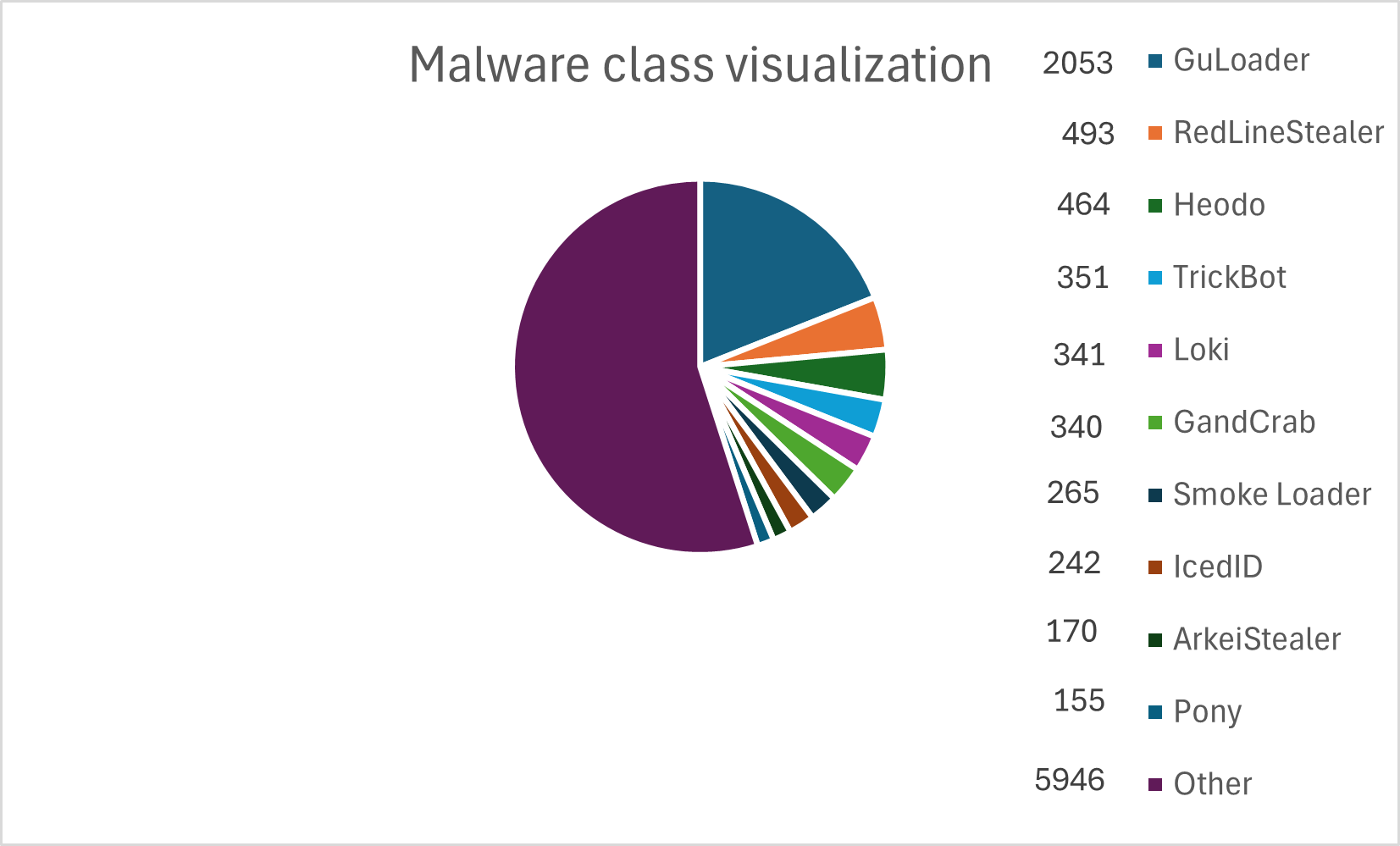}
    \caption{Malware Class Distribution}
    \label{MalwareClass}
\end{figure}

Figure~\ref{MalwareClass} presents the malware family distribution by sample count. For clarity, the ten most prevalent families are shown individually, while the remaining families are grouped under a single category. The dataset encompasses a wide range of malware behaviors, including loaders (e.g., GuLoader, SmokeLoader), information stealers (e.g., RedLineStealer, ArkeiStealer, Loki, Pony), modular banking trojans (e.g., TrickBot, IcedID, Heodo), and ransomware families such as GandCrab. This diversity, together with the presence of over 200 additional families, ensures substantial heterogeneity for evaluating anomaly detection performance.

Malicious samples were split evenly between validation and testing, with 5,410 samples assigned to each. To further assess transferability, we evaluate the trained model on the Microsoft Malware Challenge dataset~\cite{kaggle2015}, which contains 10,868 samples from nine malware families collected in 2015. This dataset serves as an external benchmark for evaluating generalization to previously unseen malware distributions.

\subsubsection{Extended Primary Dataset}

To examine the effect of increased benign data diversity, we extend the benign training set using samples from Michael Lester’s curated PE dataset~\cite{lester2021pe}. Only benign files were incorporated, and the validation and test sets remained unchanged to isolate the effect of training data expansion. The benign training set was increased from 3,798 to 15,000 samples by randomly selecting 11,202 additional benign files using a fixed seed of 42.

\subsubsection{Secondary Dataset (Contagio)}

A secondary experiment focuses on Object Linking and Embedding (OLE) files, specifically PDFs, using the Contagio dataset. This dataset includes 9,000 benign and 10,980 malicious PDF files, enabling evaluation across a different binary format and providing insight into cross file type generalization.

\subsection{Conversion of Raw Files to Features}

Feature extraction is performed directly on PE and OLE files, though the approach is applicable to other binary formats. Recent literature surveys~\cite{ngo2020survey,sihwail2018survey,pan2020systematic,vu2019hit4mal} were consulted to identify commonly used features and data representations for malware anomaly detection. Byte sequences and opcodes were selected due to their prevalence and compatibility with image based modeling.

\subsection{Feature Selection and Extraction}

\subsubsection{Byte Sequences and Opcode Identification}

Byte sequences capture raw hexadecimal content, while opcodes represent executable instructions. These features provide differentiating views of program structure and behavior for us to test.

\subsubsection{Automation with Linux Terminal Commands}

Extraction was automated using Linux command line tools integrated into Python scripts. Hexadecimal dumps were generated using \texttt{xxd}, while opcode extraction relied on \texttt{objdump}, followed by custom parsing to isolate opcode sequences.

\subsection{Conversion of Extracted Features into Images}

The extracted features are converted into images through a combination of pixel color assignment and pixel mapping. Traditional approaches, such as that of Nataraj et al.~\cite{nataraj2011malware}, map bytes line by line into grayscale images of fixed width.

Vu et al.~\cite{vu2019hit4mal} proposed the use of space filling curves to better preserve local structure. Motivated by these findings, we adopt the Hilbert curve for pixel mapping, which maintains locality during the transformation from one dimensional sequences to two dimensional grids. Prior work by Saridou et al.~\cite{saridou2023image} demonstrated the effectiveness of this approach in malware classification.

\subsubsection{Pixel Color Allocation}

Color allocation follows a standardized scheme designed to maximize geometric contrast in RGB space. Building on the ideas of Saridou et al.~\cite{saridou2022sagmad}, hexadecimal values are mapped to RGB vectors chosen to be maximally distant within the geometric color cube, enhancing discriminability in reconstruction based anomaly detection. The resulting mapping is summarized in Table~\ref{collorallocation}.

\begin{table}[h!]
\centering
\begin{tabular}{| >{\centering\arraybackslash}m{2.2cm} | c | c | c |}
\hline
\textbf{Hexadecimal Character} & \textbf{R value} & \textbf{G value} & \textbf{B value} \\
\hline
0 & 0 & 0 & 0 \\
1 & 128 & 0 & 0 \\
2 & 154 & 99 & 36 \\
3 & 128 & 128 & 0 \\
4 & 70 & 153 & 144 \\
5 & 0 & 0 & 117 \\
6 & 230 & 25 & 75 \\
7 & 245 & 130 & 49 \\
8 & 255 & 225 & 25 \\
9 & 191 & 239 & 69 \\
A & 60 & 180 & 75 \\
B & 66 & 212 & 244 \\
C & 67 & 99 & 216 \\
D & 145 & 30 & 180 \\
E & 240 & 50 & 230 \\
F & 255 & 255 & 255 \\
\hline
\end{tabular}
\caption{Standardized color allocation to hexadecimal values}
\label{collorallocation}
\end{table}

\subsubsection{Pixel Mapping}

\begin{figure}
    \centering
    \includegraphics[width=0.5\linewidth]{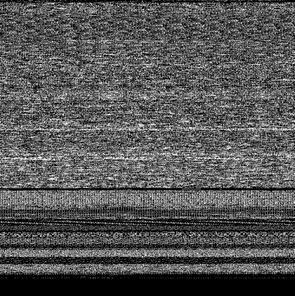}
    \caption{Malware sample converted using Nataraj et al. method}
    \label{natarajmethod}
    \vspace{-3mm}
\end{figure}

\begin{figure}
    \centering
    \includegraphics[width=0.5\linewidth]{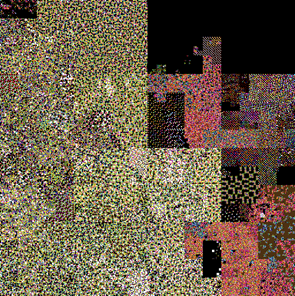}
    \caption{Malware sample converted using our method}
    \label{ourimagemethod}
    \vspace{-4mm}
\end{figure}

\begin{figure}
    \centering
    \includegraphics[width=1\linewidth]{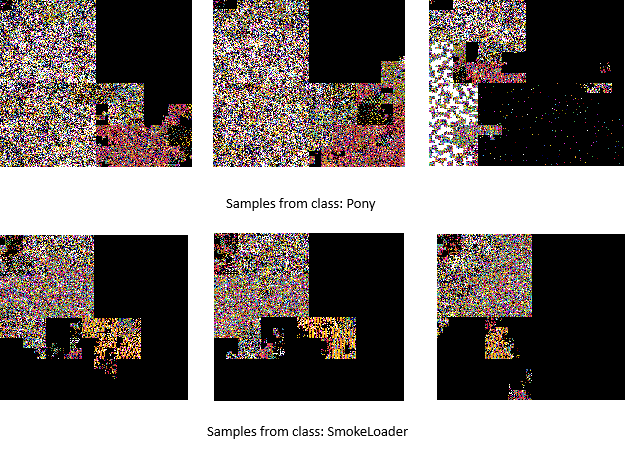}
    \caption{Malware Visualization Class Variation}
    \label{MalwareClassVis}
\end{figure}

Pixel mapping is performed using the Hilbert curve to preserve locality. Figures~\ref{natarajmethod} and~\ref{ourimagemethod} compare the conventional linear mapping with the proposed method, while Figure~\ref{MalwareClassVis} illustrates intra and interclass visual patterns.

Hilbert transformations were implemented in Python, with processing times of approximately 6 seconds for 50\,MB files and an average of 0.2 seconds per sample across the dataset, aided by caching of transformation indices.

\subsection{Model}

The proposed model builds on the CBiGAN framework of Carrara et al.~\cite{carrara2021combining}, which extends the BiGAN architecture~\cite{donahue2016adversarial} by enforcing consistency between encoded and reconstructed representations. To better accommodate the complexity of malware images, the original encoder is replaced with convolutional neural networks commonly used in image analysis, including ResNet50, ResNet101, ResNet152, DenseNet169, and InceptionV3.

To maintain architectural balance, corresponding modifications are applied to the generator. This allows exploration of trade offs between computational efficiency and anomaly detection performance. While deeper encoders increase training time, inference latency on trained models remains negligible, enabling practical deployment scenarios.

\section{Experimental Setup}

All experiments were conducted on an Ubuntu Linux workstation equipped with an Intel i9-11900K CPU, 128\,GB system RAM, and an NVIDIA RTX 3090 GPU (24\,GB VRAM).

\subsection{Evaluation Metrics}

We report Area Under the ROC Curve (AUC) and balanced accuracy as primary metrics. These were selected because standard accuracy can be misleading under class imbalance, which is common in malware screening scenarios. Balanced accuracy measures performance symmetrically across the benign and malicious classes:
\[
\text{BalAcc} = \frac{1}{2} \left( \frac{TP}{TP + FN} + \frac{TN}{TN + FP} \right),
\]
where $TP$, $FN$, $TN$, and $FP$ denote true positives, false negatives, true negatives, and false positives, respectively.

AUC summarizes separability across decision thresholds and is computed as the integral of the true positive rate (TPR) with respect to the false positive rate (FPR):
\[
\text{AUC} = \int_{0}^{1} \text{TPR}(\text{FPR}) \, d(\text{FPR}),
\]
where $\text{TPR} = \frac{TP}{TP+FN}$ and $\text{FPR} = \frac{FP}{FP+TN}$.

In addition, we report the following diagnostics to better characterize operational behavior:

\textbf{AUCPR.} We report the area under the Precision--Recall curve (malicious as the positive class), which is often more informative under heavy class imbalance.

\textbf{FPR@TPR.} We report the false positive rate at fixed true positive rates of 95\% and 99\% to quantify performance in high-recall operating regimes relevant to malware screening.

\textbf{Threshold sensitivity.} To assess operating-point stability, we sweep decision thresholds over the score range and summarize the corresponding changes in TPR, FPR, and balanced accuracy.

\textbf{Calibration error.} We measure Expected Calibration Error (ECE; 10 bins) and visualize reliability diagrams to assess the alignment between predicted confidence and empirical accuracy.

\section{Initial Experiments and Design Choice Validation}

Before running the full set of evaluations, we performed a set of controlled experiments to validate key design choices in the pipeline. These experiments are best interpreted as ablations. Each test isolates a single component (feature representation, image mapping, or encoder choice) while holding the remaining configuration fixed. Unless stated otherwise, models were trained on 3,798 benign samples and evaluated using 1,266 benign samples alongside 5,410 malicious samples for validation and 5,410 malicious samples for testing.

\subsection{Byte Sequence vs. Opcode Representation}

\subsubsection{Byte Sequence Method}
In the byte sequence setting, PE files are converted into images directly from raw byte content. This representation preserves global structural patterns and serves as input to a CBiGAN--ResNet50 configuration. Under this setting, the model achieved an AUC of 0.816.

\subsubsection{Opcode Method}
In the opcode setting, images are constructed from opcode sequences extracted from executables, aiming to emphasize instruction-level behavior. In our experiments, this approach produced a lower AUC of 0.616. A likely explanation is that opcode only representations discard portions of the file structure that remain informative under image based anomaly detection, particularly when the scoring mechanism is driven by reconstruction discrepancies.

Based on these results, the byte sequence representation was adopted for subsequent experiments.

\subsection{Effect of Hilbert Mapping and RGB Color Allocation}

We next evaluated whether locality preserving mapping and standardized RGB allocation improve anomaly detection compared to conventional byteplot rendering. For controlled comparison, we used the Contagio dataset as it offers a more constrained PDF-focused setting for early validation. Using identical data splits and the same GAN--encoder configuration, Table~\ref{initialPDFtests} shows that locality preserving mapping combined with RGB allocation substantially improves performance relative to the traditional line by line approach of Nataraj et al.~\cite{nataraj2011malware}. The improvements suggest that preserving local neighborhood relationships and increasing channel expressivity both benefit reconstruction based scoring.

\begin{table*}[h!]
\centering
\begin{tabular}{| c | c | c |}
\hline
\textbf{Method} & \textbf{AUC} & \textbf{Balanced Accuracy} \\
\hline
Traditional method \cite{nataraj2011malware} & 0.59 & 0.627  \\
\hline
Greyscale + Hilbert & 0.786 & 0.734 \\
\hline
\textbf{Our method (RGB + Hilbert)} & \textbf{0.795} & \textbf{0.782} \\
\hline
\end{tabular}
\caption{Comparison of image processing methods between Nataraj et al.\ and our approach using the Contagio dataset.}
\label{initialPDFtests}
\end{table*}

\subsection{Encoder Replacement in CBiGAN}

A key hypothesis in this work is that replacing the base CBiGAN encoder with a deep convolutional encoder improves anomaly detection performance on malware imagery. To test this under a controlled setting, we compared models trained with identical configurations and data, varying only the encoder. Table~\ref{encodertest} summarizes test and validation performance across encoders.

Replacing the base encoder yields a substantial improvement; for example, ResNet50 improves test AUC from 0.667 to 0.816. Interestingly, deeper encoders do not consistently outperform lighter ones. This may reflect the reconstruction driven nature of the anomaly score. Capturing overly rich features can reduce sensitivity to the structured color/layout discrepancies emphasized by our visualization scheme. We treat this observation as an empirical trade-off rather than a universal claim, noting that further tuning could alter this relationship.

\begin{table}
\centering
\resizebox{\columnwidth}{!}{%
\begin{tabular}{| >{\centering\arraybackslash}m{3cm} | >{\centering\arraybackslash}m{1cm} | >{\centering\arraybackslash}m{1cm} | >{\centering\arraybackslash}m{1cm} | >{\centering\arraybackslash}m{1cm} |}
\hline
\textbf{Encoder model used} & \textbf{Test AUC} & \textbf{Test Bal. Acc.} & \textbf{Valid. AUC} & \textbf{Valid. Bal. Acc.} \\
\hline
CBiGAN base encoder & 0.667 & 0.662 & 0.661 & 0.659 \\
\hline
\textbf{ResNet50} & \textbf{0.816} & \textbf{0.781} & \textbf{0.818} & \textbf{0.787} \\
\hline
ResNet101 &  0.766 & 0.731  & 0.768  &  0.730 \\
\hline
ResNet152 & 0.791 & 0.737 & 0.778 & 0.724 \\
\hline
DenseNet169 & 0.801 & 0.730  & 0.806 & 0.736  \\
\hline
InceptionV3  & 0.833  & 0.762  &  0.620 & 0.599  \\
\hline
\end{tabular}
}
\caption{Test and validation performance using different encoders on the self-collected PE dataset (214 families).}
\label{encodertest}
\end{table}

\subsection{Effect of Training Set Size}

Finally, we evaluated whether increasing benign training diversity improves performance while keeping the validation and test sets unchanged. As shown in Table~\ref{datasizetest}, increasing the benign training set from 3,798 to 15,000 improves both AUC and balanced accuracy, supporting the hypothesis that additional benign variation helps the one class model better characterize normal structure and reduce false positives.

\begin{table}
\centering
\resizebox{\columnwidth}{!}{%
\begin{tabular}{| >{\centering\arraybackslash}m{3cm} | >{\centering\arraybackslash}m{1cm} | >{\centering\arraybackslash}m{1cm} | >{\centering\arraybackslash}m{1cm} | >{\centering\arraybackslash}m{1cm}|}
\hline
\textbf{Training Dataset Size} & \textbf{Test AUC} & \textbf{Test Bal. Acc.} & \textbf{Valid. AUC} & \textbf{Valid. Bal. Acc.} \\
\hline
3,798 & 0.816 & 0.781 & 0.818 & 0.787 \\
\hline
\textbf{15,000} & \textbf{0.864} & \textbf{0.831} & \textbf{0.857} & \textbf{0.822} \\
\hline
\end{tabular}
}
\caption{Test and validation results for CBiGAN--ResNet50 under different benign training set sizes.}
\label{datasizetest}
\vspace{-3mm}
\end{table}

\begin{table*}[h!]
\centering
\resizebox{\textwidth}{!}{%
\begin{tabular}{| >{\centering\arraybackslash}m{1.9cm} | >{\centering\arraybackslash}m{1cm} | >{\centering\arraybackslash}m{1.1cm} | >{\centering\arraybackslash}m{1cm} | >{\centering\arraybackslash}m{1.4cm} | >{\centering\arraybackslash}m{1cm} | >{\centering\arraybackslash}m{1cm} | >{\centering\arraybackslash}m{1cm} | >{\centering\arraybackslash}m{1.3cm} | >{\centering\arraybackslash}m{1.4cm} | >{\centering\arraybackslash}m{0.9cm} |}
\hline
\textbf{Name} & \textbf{Best AUC} & \textbf{MS1 - ramnit }& \textbf{MS2 - Lollipop }& \textbf{MS3 -- kelihos\_ver3} & \textbf{MS4 - Vundo} & \textbf{MS5 - Simda} & \textbf{MS6 - Tracur} & \textbf{MS7 -- Kelihos\_ver1} & \textbf{MS8 -- Obfuscator.ACY} & \textbf{MS9 - Gatak} \\
\hline
CBiGAN-D169 & 0.801 & 0.555 & 0.448 & 0.252 & 0.353 & \textbf{0.771} & 0.456 & 0.916 & 0.478 & 0.568 \\
\hline
CBiGAN--R152 & 0.791 & 0.629 & \textbf{0.493} & 0.149 & 0.279 & 0.097 & 0.41 & 0.887 & 0.823 & 0.539 \\
\hline
CBiGAN--R50 & 0.816 & 0.603 & 0.419 & \textbf{0.357} & 0.218 & 0.328 & \textbf{0.513} & 0.852 & 0.842 & 0.396 \\
\hline
CBiGAN--R50 (larger training set) & 0.864 & \textbf{0.683} & 0.413 & 0.314 & \textbf{0.724} & 0.618 & 0.475 & \textbf{0.925} & \textbf{0.886} & \textbf{0.665} \\
\hline
\end{tabular}
}
\caption{Transferability of the trained models to unseen malware families in the Microsoft Malware Challenge dataset.}
\label{transferability}
\end{table*}

\section{Experiments}

This section presents a set of controlled experiments designed to systematically evaluate the impact of key design choices in the proposed CBiGAN based anomaly detection framework. These experiments are structured as ablation studies, isolating the influence of encoder architecture, training data diversity, and file-type variation on detection performance and generalization. The goal is not to introduce new modeling components, but to quantify how each design decision affects robustness and transferability in realistic malware detection scenarios.

Across all experiments, models are trained exclusively on benign samples and evaluated using balanced accuracy and AUC. In addition to in-distribution testing, we assess zero-shot generalization by evaluating trained models on previously unseen malware families, which is a critical requirement in
operational cybersecurity settings where novel threats continually emerge.

During training, two checkpoints are maintained: one corresponding to the model state achieving the highest AUC on the test set, and a second capturing the final model state after training completion. This allows comparison between peak
separability performance and end of training behavior. An exponential moving average (EMA) of model parameters is applied to stabilize training dynamics and reduce metric volatility, yielding more reliable estimates of model performance.

\subsection{Ablation Study: Baseline Performance on PE Files}

The baseline experiment evaluates the proposed method on PE files using the self-collected primary dataset. This experiment establishes a reference point for subsequent ablations and focuses on assessing performance across a large and heterogeneous set of malware families. By incorporating samples from 214 distinct families, this setup emphasizes generality rather than specialization to a narrow threat class.

\subsection{Ablation Study: Effect of Benign Training Set Size}

To examine the role of benign data diversity, we evaluate the impact of expanding the benign training set while keeping all other variables fixed. Specifically, the original training set of 3,798 benign samples is extended to 15,000 samples using additional benign executables from Michael Lester’s curated dataset \cite{lester2021pe}. This experiment isolates the effect of training data scale on anomaly detection performance and transferability, using a ResNet50-based
CBiGAN configuration with identical hyperparameters.

\subsection{Ablation Study: Zero-Shot Transferability Across Malware Families}

This experiment evaluates the model’s ability to generalize to previously unseen malware families in a zero-shot setting. Models trained on the primary PE dataset are tested against the Microsoft Malware Challenge dataset \cite{kaggle2015}, which contains nine malware families not present during training. This setting simulates real world deployment conditions, where detectors must identify novel threats without prior exposure.

\subsection{Ablation Study: Cross Format Generalization to OLE/PDF Files}

Finally, we assess cross-format robustness by evaluating the proposed approach on
OLE files, specifically PDF documents from the Contagio dataset. This experiment
examines whether a model trained using the same visualization and reconstruction
principles can effectively operate in a different binary context, providing
insight into the broader applicability of the method beyond PE executables.

\section{Results}

\subsection{Primary Experiment Results}
\begin{figure}
    \centering
    \includegraphics[width=0.75\linewidth]{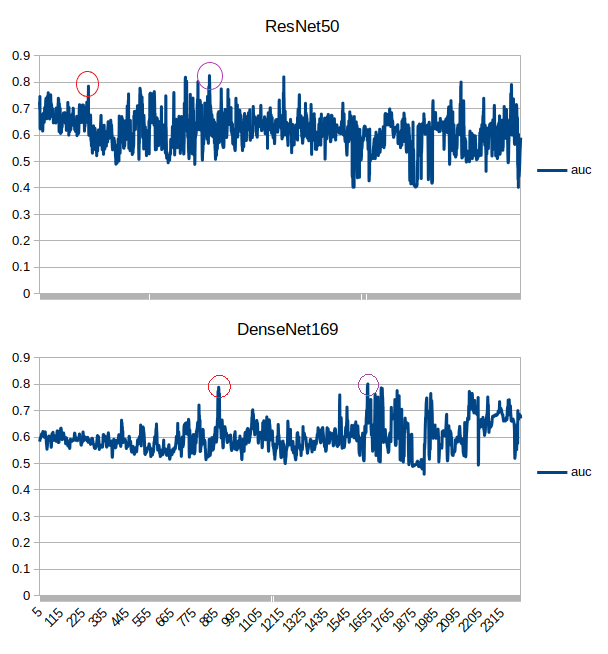}
    
    \caption{Comparison of AUC over time for ResNet50 and DenseNet169 encoders}
    \label{AUCCompariosn}
    \vspace{-3mm}
    
\end{figure}

The comparative analysis of the performance across ResNet50, ResNet152, and DenseNet169 models as shown in table \ref{encodertest}, provides a nuanced view of the effectiveness of different network architectures in our malware detection framework. While ResNet50 achieves a higher AUC in both test and validation sets, and reaches its performance peaks earlier in the training process, this does not necessarily translate to superior generalizability across all scenarios, notably, the ResNet models reach their peak AUC considerably earlier than the DenseNet model as observable in figure \ref{AUCCompariosn} which shows the comparison of AUC over time for the ResNet50 an DenseNet169 models. Initially, this might suggest a preference for ResNet models, however, a deeper examination of model transferability reveals a different perspective. 

ResNet50’s earlier peaking in AUC might indicate faster convergence, potentially due to its shallower architecture compared to DenseNet169. This can be advantageous in scenarios where quick deployment of updated models is crucial. However, the lower peak AUCs of ResNet152 and DenseNet169 suggest these models may capture more complex patterns that are not immediately evident early in training.

\subsection{Dataset Size Experiment Results}
Table \ref{datasizetest} shows the results of training an identical model to the CBiGAN-ResNet50 model, which showed the best results in the prior experiment, with an increased training dataset. We tested by increasing the dataset size from 3,798 to 15,000 by supplementing with benign samples from Michael Lester's PE machine learning dataset \cite{lester2021pe}. Samples were selected based on a seed of 42. As we can see there is a positive correlation in increasing the dataset size and variation of samples in this context with an improved AUC of 0.864 as opposed to the 0.816 of the smaller training set. This passes on to the validation set as well with an AUC of 0.857 as opposed to 0.818 of the smaller training set.

\subsection{Transferability Experiment Results}
Table \ref{transferability} further provides a critical insight into the transferability of the models, by testing the models trained on our self-collected dataset against the 9 classes of the Microsoft malware challenge dataset\cite{kaggle2015}. Despite their slower ascent to peak performance, both ResNet152 and DenseNet169 demonstrate a reasonable capability in generalizing to the Microsoft malware classification challenge dataset, including various malware types unseen during training. This indicates that the depth of the model plays a reasonable role in learning generalized features rather than fitting them to the specificities of the training data.

The ability of deeper models to perform better on a comprehensive challenge such as the Microsoft dataset, despite lower AUCs initially, underlines the importance of architectural depth for complex image recognition tasks like malware detection. While ResNet50 may be efficient for quicker tasks and performs well on known test and validation sets, models like ResNet152 and DenseNet169 are more robust when dealing with diverse and previously unseen malware classes. Therefore, while ResNet50 provides a good balance between performance and speed, for applications where robustness against diverse threats is critical, deeper networks might offer more reliable protection, albeit at the cost of increased computational resources and training time.

However, the most influential factor for transferability results was the increased dataset size. One could conclude that the increased variation in the training set would affect the transferability results positively. Although the results were not great through each of the 9 classes in the Microsoft dataset, we do see a clear increase in the results of the transferability test, with the increased dataset size test showing the highest result for 5 of 9 classes tested on. It also shows a relatively positive result for MS5 - Simda, which for the other ResNet results did not show good results. We can also see a correlation in the encoder model used and results for certain classes, such as for MS8 - Obfuscator.ACY showing subpar results for the DenseNet experiment and >0.8 AUC results for the ResNet models.
\begin{table}[]
\centering

\begin{tabular}{| c  | c  |}
\hline
\textbf{Model Used}& \textbf{AUC}\\
\hline
\textbf{ResNet50} & \textbf{0.837}\\
\hline
ResNet101 &  0.74\\
\hline
ResNet152 & 0.767\\
\hline
DenseNet169 & 0.772\\
\hline
InceptionV3  & 0.766\\
\hline

\end{tabular}

\caption{Results using different encoders  against Contagio dataset}
\label{secondaryexperiment}
\vspace{-4mm}
\end{table}

\subsection{Secondary Experiment Results}

For the secondary experiment we tested our model against the Contagio PDF dataset\cite{mila2013dataset}. Table \ref{secondaryexperiment} shows the results of using different encoder variants with the Contagio dataset, with the ResNet50 encoder showing the best results by far with 0.837 AUC score.
We also test the usage of limiting the PDF files to only ones containing JavaScript components to see the effect on the AUC score, based on the findings of Gu et al. \cite{gu2023pdfdetection}. For this we utilize the full dataset as well as follow the same method as Gu et al. by using PeePDF to separate the samples with JavaScript components. 

\section{Analysis}
\subsection{Evaluation of Best Performing Model}
To evaluate our best performing model; CBiGAN-R50 with increased dataset size, we generated the following.

\subsubsection{ROC Curve}
\begin{figure}
    \centering
    \includegraphics[width=1\linewidth]{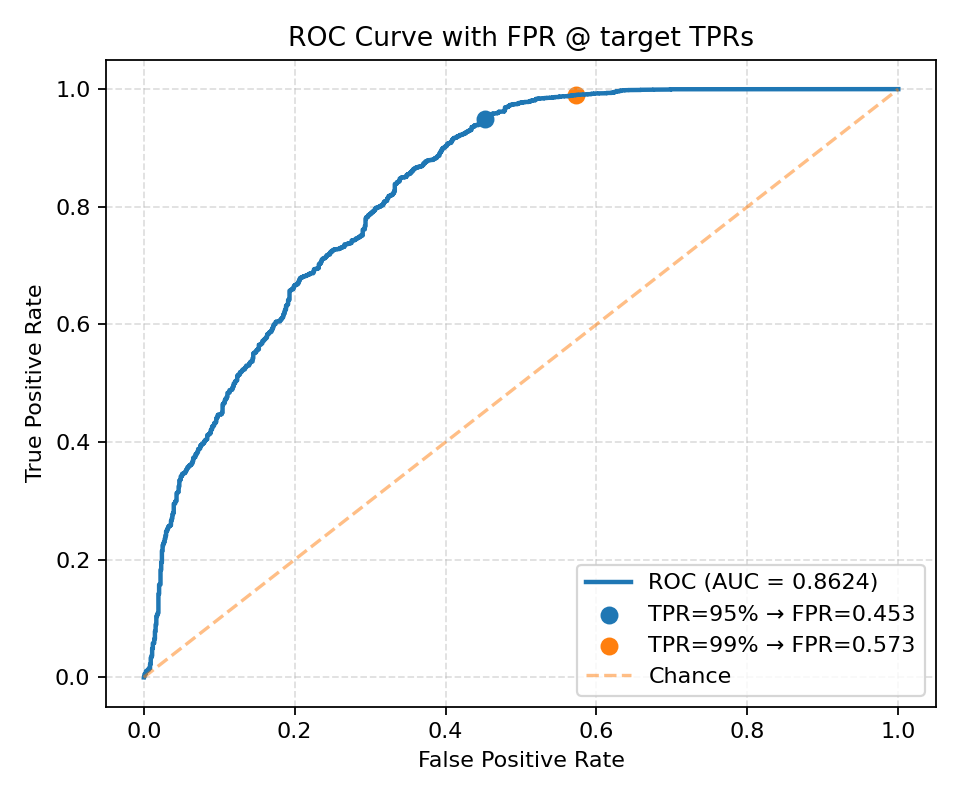}
    \caption{ROC graph with markers for FPR at TPR=99/95}
    \label{ROCmarker}
\end{figure}
Figure \ref{ROCmarker} essentially shows how well a model distinguishes between benign and malicious samples across different decision thresholds. The overall curve shows that the model has learned useful discriminative patterns. It shows a gradual rise on the left side of the graph, which shows that at the right threshold it can achieve reasonably high detection rates with low false positives. However if you consider the markers which show the points at which TPR = 95\% and 99\%, it shows a somewhat high FPR. This shows that at the right threshold however we can achieve a good TPR value with acceptable FPR.

\subsubsection{PR Curve}
\begin{figure}
    \centering
    \includegraphics[width=1\linewidth]{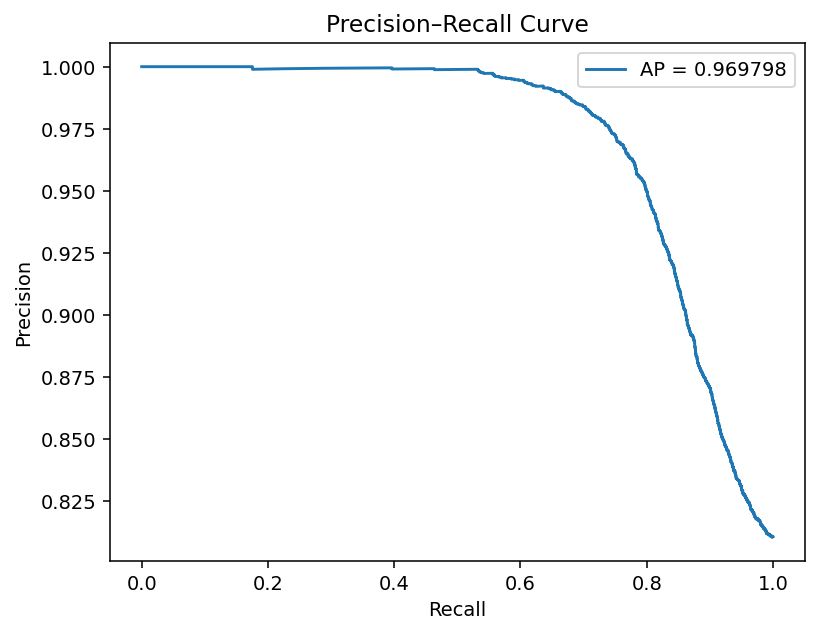}
    \caption{AUCPR graph}
    \label{AUCPR}
\end{figure}

Figure \ref{AUCPR} demonstrates consistently high precision across most recall levels, with an average precision (AP) of 0.97. This indicates that the model maintains a very low false positive rate until extremely high recall, where precision gradually declines.

\subsubsection{Threshold Sensitivity}
\begin{figure}
    \centering
    \includegraphics[width=1\linewidth]{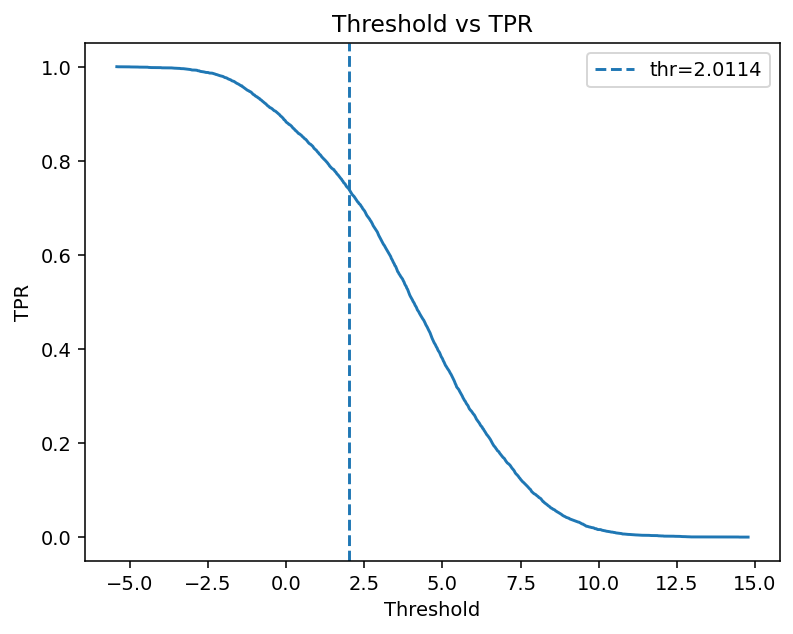}
    \caption{Threshold vs TPR}
    \label{TPR}
\end{figure}

\begin{figure}
    \centering
    \includegraphics[width=1\linewidth]{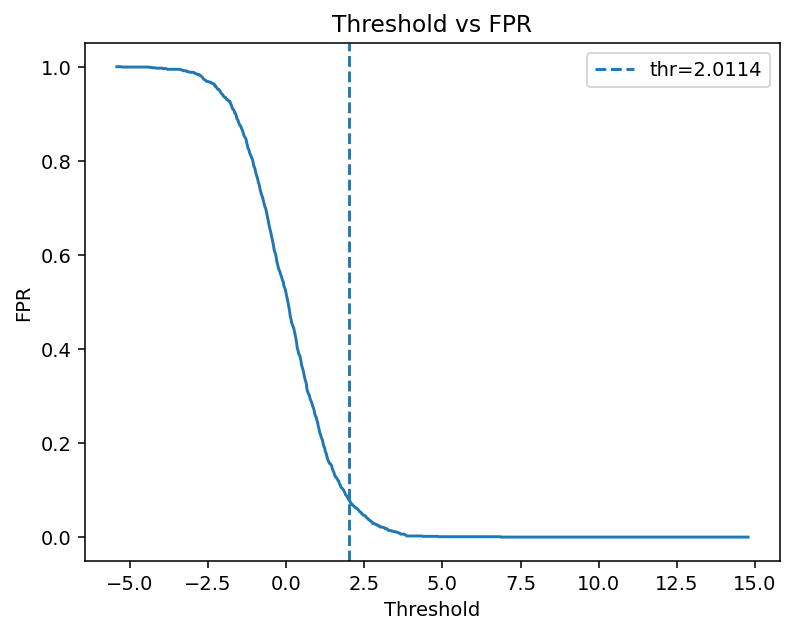}
    \caption{Threshold vs FPR}
    \label{FPR}
\end{figure}

\begin{figure}
    \centering
    \includegraphics[width=1\linewidth]{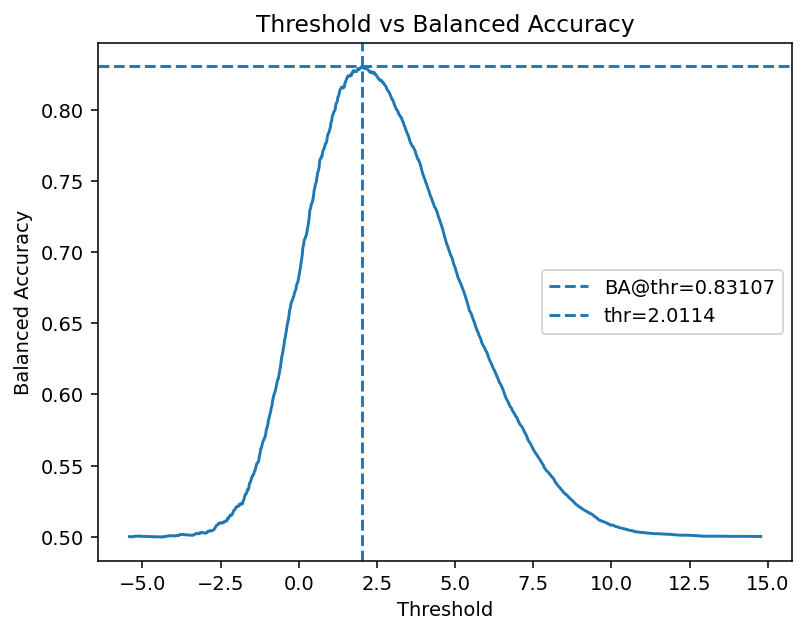}
    \caption{Threshold vs Balanced accuracy}
    \label{ThrBalacc}
\end{figure}

Figures \ref{TPR}, \ref{FPR} and \ref{ThrBalacc} illustrate how varying the decision threshold impacts performance. As the threshold increases, the false positive rate (FPR) decreases while the true positive rate (TPR) also drops, highlighting the tradeoff between sensitivity and specificity. The balanced accuracy curve shows an optimal threshold around 2.0, where the model achieves its maximum balanced accuracy of 0.83.

\subsection{Calibration Performance}
\begin{figure}
    \centering
    \includegraphics[width=1\linewidth]{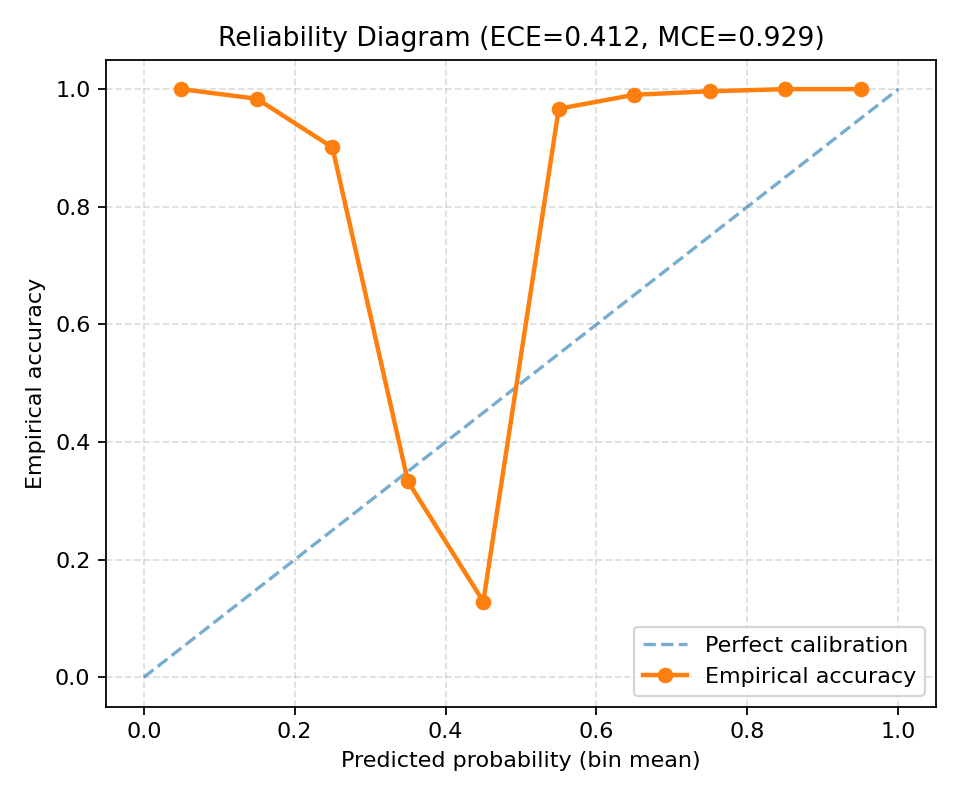}
    \caption{Calibration reliability diagram}
    \label{calibration}
\end{figure}
The reliability diagram in Fig.\ref{calibration} demonstrates substantial miscalibration between the model’s predicted probabilities and the corresponding empirical accuracies. The ideal calibration line (dashed) indicates perfect confidence alignment, while the orange curve shows that the model performs well at extreme probability bins but exhibits notable deviation in the mid confidence range (around 0.3–0.5). This “U-shaped” trend suggests that the model is overconfident in uncertain predictions and underestimates reliability in intermediate regions, resulting in inconsistent confidence estimates. The high Expected Calibration Error (ECE = 0.412) and Maximum Calibration Error (MCE = 0.929) further confirm that although the classifier achieves high accuracy for confident predictions, calibration varies across the full probability spectrum.

The observed miscalibration highlights the importance of post-hoc calibration and threshold optimization in anomaly detection systems, particularly in high recall operational regimes.

\subsection{Statistical Validation and Benchmarks}
\subsubsection{Statistical Validation}
To assess result stability and reproducibility, we conduct multi-seed evaluation and report mean and standard deviation across independent runs.

To further test the consistency of our method, we reran our best performing model (i.e., the CBiGAN-R50 with increased dataset) using two additional random seeds capped at 99. Our initial tests were all performed with the set seed of 42.

\begin{table}[h!]
\centering
\begin{tabular}{|c|c|c|}
\hline
\textbf{Seed} & \textbf{AUC} & \textbf{Balanced Accuracy} \\
\hline
42 & 0.864 & 0.831 \\
14 & 0.855 & 0.819 \\
93 & 0.861 & 0.821 \\
\hline
\textbf{Mean $\pm$ Std} & 0.860 $\pm$ 0.0046 & 0.824 $\pm$ 0.0064 \\
\hline
\end{tabular}
\caption{Statistical validation of the CBiGAN-ResNet50 model with increased dataset across three seeds.}
\label{statval}
\end{table}

Table \ref{statval} shows that our initial test run at seed 42 showcases the best result, however the additional two random seeds still show consistent and comparable results with acceptable standard deviation values. While formal hypothesis testing is limited by dataset reuse constraints, the low variance across seeds indicates consistent and stable model behavior.

\begin{table}[h!]
\centering
\resizebox{\columnwidth}{!}{%
\begin{tabular}{|l|c|}
\hline
\textbf{Process} & \textbf{Time Taken} \\
\hline
Average image transformation & $\approx$ 0.02 s \\
\hline
Image transformation (smallest sample, 2 MB) & $<$ 0.01 s \\
\hline
Image transformation (largest sample, 50 MB) & $\approx$ 6 s \\
\hline
Time to decision on trained model & $<$ 0.01 s \\
\hline
\end{tabular}%
}
\caption{Average and case-specific time measurements for image transformation and model inference.}
\label{timings}
\end{table}

The results in Table \ref{timings} were obtained on a system with an Intel i9-11900K CPU, 128 GB of RAM, and an NVIDIA RTX 3090 GPU (24 GB VRAM) utilizing CUDA acceleration. These measurements indicate that image transformation is lightweight on average but scales with file size, while inference on the trained model remains near instantaneous.

\subsection{Comparison with Shaukat et al. \cite{shaukat2024detection}}
Our principal comparison for the PE dataset is with the model proposed by Shaukat et al., which employs a one-class SVM trained solely on benign PE files converted to RGB images using a method adapted from Nataraj et al. Shaukat et al.'s study is one of the few utilizing malware images for anomaly detection rather than classification. They test their initial model against a combination of the Microsoft malware challenge dataset and VirusShare. They show their results for their model alone as well as with an extra feature extraction process using PCA and a deep learning model. Table \ref{shaukatres} is an extract from Shaukat et al.'s paper which shows their key results.

\begin{table}
\centering
\begin{tabular}{| c | >{\centering\arraybackslash}m{2cm}  | c  |}
\hline
 \textbf{Model Used}& \textbf{Feature Reduction Step}& \textbf{AUC}\\
\hline
 OSVM + RegNetY320& No& 0.89\\
\hline
 OSVM + RegNetY320& Yes& 0.921\\
\hline
  OSVM + ResNet152& No&   0.84\\
\hline
 OSVM + ResNet152& Yes& 0.87\\
\hline
 OSVM + VGG19& No& 0.81\\
\hline
 OSVM + VGG19& Yes&  0.891\\\hline
\end{tabular}
\caption{Results of Shaukat et al.}
\label{shaukatres}
\vspace{-3mm}
\end{table}

Our CBiGAN-ResNet50 model demonstrates comparable results with an AUC of 0.816 on the PE dataset, which contains over 214 malware classes as opposed to the 9 classes used by Shaukat et al. Furthermore, we tested their feature reduction technique within our pipeline, which increased the CBiGAN-ResNet50 AUC to 0.837 but negatively affected transferability for several Microsoft malware classes (notably classes 5, 7, and 8). This highlights a trade-off between in-distribution performance and cross-family generalization.

It is important to note that these methods operate under different threat models and design assumptions; therefore, comparisons are intended to contextualize design trade-offs rather than assert direct superiority.

\subsection{Comparison with Gu et al. \cite{gu2023pdfdetection}}
Our analysis also compares our methodology against the JavaScript-based detection approach proposed by Gu et al., which focuses on JavaScript code within PDF documents. We propose this comparison as both methods utilize the publicly available Contagio OLE dataset \cite{mila2013dataset}. Gu et al. extract and tokenize JavaScript using PJscan and peepdf parsers, followed by one-class SVM classification.

Gu et al. report that 97.51\% of malicious PDF files in their dataset contain JavaScript, compared to only 4.26\% of benign files, achieving a detection accuracy of 96.9\% with a false positive rate of 3.2\%. Our results are lower, with an AUC of 0.837 using the ResNet50 encoder. Restricting evaluation to PDFs containing JavaScript marginally increased AUC to 0.855.

While their results are notable, the dependency on JavaScript presence may limit generalization as malware authors adapt to detection strategies. In contrast, our CBiGAN-based approach does not rely on explicit semantic features such as JavaScript, enabling broader applicability and robustness to obfuscation. This is further supported by our transferability results on the obfuscater.acy malware class, where an AUC of 88.6\% was achieved.

It is important to note that our approach and that of Gu et al. address different detection assumptions, and the comparison is intended to illustrate complementary strengths rather than establish direct dominance.

\section{Limitations and Future Work}

\subsection{Limitations} 
While our approach demonstrates promising results, several limitations should be acknowledged. 
First, the model shows strong overall AUC and AUCPR, yet struggles to maintain low false positive rates at very high recall thresholds (e.g., FPR above 0.4 when TPR $\geq$ 95\%), limiting its immediate applicability in high sensitivity deployment settings. 
Second, the malware family labels in our dataset are unevenly distributed and in many cases not well characterized, leaving gaps in understanding the similarities among malware classes and their proximity to benign software. 
This lack of granularity makes it difficult to assess whether high anomaly scores reflect true malicious behavior or benign outliers that share surface level patterns with malware. 
Third, by design our pipeline emphasizes simplicity and reproducibility (minimal preprocessing, straightforward image conversion, and a single GAN-based model). 
While this design achieves comparability with more complex baselines, it inevitably sacrifices some performance and robustness that could be gained from multi-stage or hybrid systems. 
Fourthly, our evaluation does not fully explore model explainability: although reconstruction errors and feature discrepancies provide anomaly scores, they do not offer interpretable insights into which regions or features of a binary most strongly contribute to detection decisions. 
In addition, per family variability on the Microsoft dataset reveals texture bias and packing sensitivity, with strong results on families such as Kelihos V1 but weaker detection on Simda or Kelihos V3, suggesting domain shift between self collected (2018–2023) and older curated samples. 
Label noise and ground truth uncertainty from public feeds (e.g., PUPs, multi label overlaps) may also confound transfer evaluations. 
Temporal drift further limits generalization, as models trained on recent threats may fail on legacy malware, while our imaging approach, though preserving locality through Hilbert mapping, imposes hand crafted assumptions that underrepresent sparse yet semantically important code regions. 
Performance also shows hyperparameter and seed sensitivity, and calibration remains imperfect at high recall thresholds. 
Moreover, we were not able to conduct a robustness evaluation against packing or evasion based perturbations. 
We emphasize that robustness to adaptive adversarial manipulation remains an open problem for all static malware detectors, including both deep learning and signature-based systems.

\subsection{Future Work} 
Future research should explore integrating richer contextual knowledge of malware families, including interfamily similarity measures and benign malicious overlap, to better calibrate anomaly detection thresholds and reduce false positives. 
Explainability is another key direction. Visual saliency maps, feature attribution techniques, or latent space disentanglement could help us understand why a sample is flagged as anomalous. 
Moreover, calibration strategies (e.g., temperature scaling, isotonic regression) and threshold optimization tuned to resource budgets could improve deployment reliability in low false positive regimes. 
Also, while our current pipeline emphasizes simplicity, future iterations could combine our imaging approach with dynamic features, ensemble methods, or hybrid static dynamic frameworks. This could help balance interpretability, computational efficiency, and detection performance. 
Beyond these, time aware evaluation and continual learning could mitigate temporal drift, while class or cluster aware thresholds may address per family variance in the Microsoft dataset.

\section{Conclusion}

In this study, we demonstrate that consistency-based generative modeling can serve as an effective and scalable anomaly detection framework when combined with lightweight visual representations of malware binaries. Specifically, we investigate the use of the consistency Bi-directional Generative Adversarial Network (CBiGAN) integrated with deep learning encoders such as ResNet and DenseNet for static malware anomaly detection. By leveraging visual representations of binary content, our approach captures structural and texture-level patterns within malware binaries while maintaining a streamlined and reproducible pipeline.

Our experimental results showed that the CBiGAN achieved strong predictive performance and reasonable generalizability across diverse datasets including PE and OLE files. This method's efficacy is highlighted by its ability to handle a diverse set of malicious executables from 214 malware families with reasonable accuracy utilizing a simplistic pipeline.
Our CBiGAN-based method offers a streamlined, single model approach that contrasts with the multiple, separate steps required in other studies. This simplicity is advantageous for practical applications, where ease of deployment and maintenance are critical. Additionally, our method's ability to detect sophisticated obfuscation techniques in malware, such as those used in the obfuscater.acy malware class, underlines its effectiveness and capability in handling real-world malware threats.

\section*{Declarations}

\subsection*{Availability of data and materials}
The datasets generated and analysed during the current study are all publicly available, except for the self-collected benign training data. 

\subsection*{Competing interests}
The authors declare that they have no competing interests.

\subsection*{Funding}
This research did not receive any specific grant from funding agencies in the public, commercial or not-for-profit sectors.

\subsection*{Authors’ contributions}
Not applicable for blind draft

\subsection*{Acknowledgements}
Not applicable

\subsection*{Authors’ information}
Not applicable.

\end{document}